# A Study of the Quark–Gluon Vertex

Jon Ivar Skullerud [a] (UKQCD Collaboration)

[a]Department of Physics and Astronomy, The University of Edinburgh, Edinburgh EH9 3JZ, Scotland

We present the first results from a study of the quark–gluon vertex function in the Landau gauge. The results are obtained for quenched QCD with an $O(a)$-improved Wilson fermion action, at $\beta = 6.0$. We discuss plans for further study, including extraction of a renormalised QCD coupling from the momentum dependence of the vertex.

## 1. Introduction

The study of quark and gluon correlation functions on the lattice is emerging as a useful tool for increasing our understanding of QCD. Studies of the gluon propagator [1,2] have given valuable insight into the phenomena of confinement and dynamical mass generation. The three-gluon vertex has been shown [3] to provide a method for determining the running coupling from first principles, and preliminary results for the quark propagator [4] indicate that this will yield useful information on chiral symmetry breaking. Here, the methods will be applied to the quark–gluon vertex.

Studying the quark–gluon vertex on the lattice gives us yet another way of determining the running coupling, similar to but independent of the determination from the 3-gluon vertex, to complement other determinations. It may also provide input to Dyson–Schwinger equation calculations, and give insight into the limits of perturbation theory by indicating where the nonperturbative lattice results deviate from perturbative calculations.

## 2. The vertex function

We can define the configuration space quark–gluon vertex function as

$$\begin{aligned}V_\mu^a(x,y,z)_{\alpha\beta}^{ij} &= \left\langle \psi_\alpha^i(x)\overline{\psi}_\beta^j(z) A_\mu^a(y)\right\rangle \\ &= \left\langle S_{\alpha\beta}^{ij}(x,z) A_\mu^a(y)\right\rangle \end{aligned} \quad (1)$$

Fourier transforming this gives us the full (unamputated) momentum space vertex function:

$$\begin{aligned}&\int d^4x\, d^4y\, d^4z\, e^{-i(px+qy-rz)} \left\langle \psi_\alpha^i(x) A_\mu^a(y)\overline{\psi}_\beta^j(z)\right\rangle \\ &= \int d^4z\, e^{-i(p+q-r)z} \int d^4x\, d^4y\, e^{-i(px+qy)} \times \\ &\quad \times \left\langle \psi_\alpha^i(x) A_\mu^a(y)\overline{\psi}_\beta^j(0)\right\rangle \\ &= (2\pi)^4 \delta(p+q-r) \left\langle S_{\alpha\beta}^{ij}(p) A_\mu^a(q)\right\rangle \\ &= (2\pi)^4 \delta(p+q-r)\, G_\mu^a(p,q)_{\alpha\beta}^{ij} \end{aligned} \quad (2)$$

and the amputated (OPI) vertex function

$$\Lambda_\mu^{a,\mathrm{lat}}(p,q) = \frac{\left\langle S(p) A_\mu^a(q)\right\rangle}{\left\langle S(p+q)\right\rangle \left\langle D(q)\right\rangle \left\langle S(p)\right\rangle} \quad (3)$$

where $D(q) = \sum_{\mu,a} D_{\mu\mu}^{aa}(q)$ is the gluon propagator.

At tree level, this becomes

$$\begin{aligned}\Lambda_\mu^{a(0)}(p,q) = &-ig_0 T^a (2\pi)^4 \times \\ &\times \left(\gamma_\mu \cos\frac{(2p+q)_\mu}{2} - ir\sin\frac{(2p+q)_\mu}{2}\right)\end{aligned} \quad (4)$$

We will also define the colour-traced full vertex function

$$G_\mu(p,q) = \frac{2}{N_C^2-1} \sum_{ij,a} T_{ij}^a G_\mu^a(p,q)_{\alpha\beta}^{ji} \quad (5)$$

In the continuum, the off-shell amputated vertex function will have the general form

$$\begin{aligned}\Lambda_\mu(p^2,q^2,pq) &\equiv \frac{1}{N_C^2-1} \mathrm{Tr}_{col} T^a \Lambda_\mu^a(p,q) \\ &= F_1 p_\mu + F_2 q_\mu + F_3 \gamma_\mu \\ &\quad + F_4\, \slashed{p} p_\mu + F_5\, \slashed{p} q_\mu + F_6\, \slashed{q} p_\mu + F_7\, \slashed{q} q_\mu \\ &\quad + F_8 \sigma_{\mu\nu} p^\nu + F_9 \sigma_{\mu\nu} q^\nu\end{aligned} \quad (6)$$

where all the $F$'s depend only on the invariants $p^2, q^2$ and $pq$.

At tree level, this reduces to $\Lambda_\mu^0 = \frac{i}{2}g_0\gamma_\mu$. From this we can see that the form factor containing the running coupling is $F_3$, while all the other form factors are expected to be finite. $F_9$ will also be of interest, since it is related to the chromomagnetic moment of the quarks.

$F_3$ may be extracted by exploiting the symmetries of the problem. First, we isolate $F_3$ and $F_7$

$$\mathrm{Tr}\,\gamma_\mu\Lambda_\mu(p\to 0,q) = 4F_3(q^2) + 4q_\mu^2 F_7(q^2) \qquad (7)$$

We can then eliminate $F_7$ by imposing an appropriate kinematics, e.g. $q = q_\nu$ with $\mu \neq \nu$.

Finally, we define the renormalised coupling as

$$g_R(q^2) = -2iZ_\psi Z_A^{1/2} F_3(q^2) \qquad (8)$$

where $Z_\psi$ and $Z_A$ are the renormalisation constants for the quark and gluon fields, defined by

$$D_{\mu\nu}(p^2)|_{p^2=q^2} = T_{\mu\nu}(q)Z_A\frac{1}{q^2} \qquad (9)$$

$$\mathrm{Tr}\,\not{p}S^{-1}(p)|_{p^2=q^2} = iZ_\psi q^2 \qquad (10)$$

where $T_{\mu\nu}$ is the projection onto transverse fields.

From this, we can compute $g_R(q^2)$ at various $\beta$ values, and relate this to the running coupling calculated in other schemes, eg, $g_R^{\overline{MS}}(q^2)$ or the lattice 3-gluon coupling [3]. In this connection, it is worth noting that the matching with $g_R^{\overline{MS}}(q^2)$ can be performed entirely within continuum perturbation theory [5].

## 3. Computation and results

In the initial study which we will report on here, we have so far looked only at the full (unamputated) vertex function, in configuration space as well as in momentum space.

36 configurations have been analysed at $\beta = 6.0$ with a lattice size of $16^3 \times 48$. The propagators have been generated using the Sheikholeslami–Wohlert action, at $\kappa = 0.1432$ (corresponding to $m_{PS}a = 0.386$). The gauge fields and propagators have been fixed to Landau gauge, with accuracy $\frac{1}{VN_c}\sum_{x,\mu}|\partial_\mu A_\mu(x)|^2 < 10^{-6}$.

For the vertex function in configuration space, the quantity $V_\mu(x_A, x_\psi) = \mathrm{Tr}_{col}\lambda^a V_\mu^a(0, x_A, x_\psi)$ has been calculated. A clear signal has been found for $\mathrm{Tr}\,V_\mu$ and $\mathrm{Tr}\,\gamma_\mu V_\mu$. The latter is shown as a function of $x_\psi$ and $x_A$ in fig. 1.

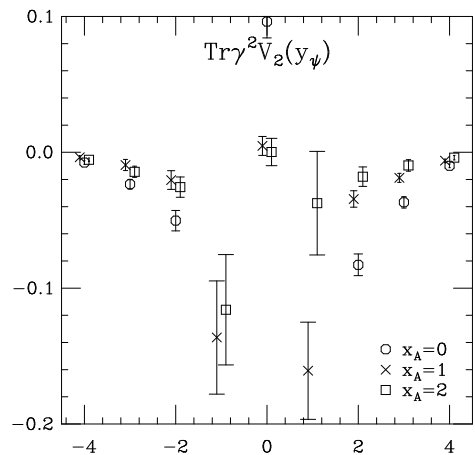

Figure 1. Vertex function as a function of $x_\psi = (0, y_\psi, 0, 0)$ and different values of $x_A = (x_A, 0, 0, 0)$, for 36 configurations, at $\kappa = 0.1432$.

We have also calculated the 'timesliced' vertex functions

$$G_\mu(\vec{p},\vec{q},t,t_A) = \mathrm{Tr}_{col}\lambda^a \int d^3x d^3y e^{-i(\vec{p}\vec{x}+\vec{q}\vec{y})} \times$$
$$\times \left\langle S(\vec{x},t;0)A_\mu^a(\vec{y},t_A)\right\rangle \qquad (11)$$

and

$$G_\mu(\vec{p},q,t) = \int dt_A e^{iq_t t_A} G_\mu(\vec{p},\vec{q},t,t_A) \qquad (12)$$

These both fall off exponentially with $t$ (after antisymmetry in $t$ has been imposed), as one should expect. Fig. 2 shows $\gamma_\mu G_\mu(\vec{p},\vec{q},t,t_A)$ as a function of $t$ for different values of $t_A$, while fig. 3 shows $\gamma_\mu G_\mu(\vec{p},q,t)$ as a function of $t$ for different values of the momenta.

$\mathrm{Tr}\,\gamma_\mu G_\mu(p,q)$ is shown as a function of $p = (0,0,0,p_t)$ for $q = 0$ (with $\mu = 3$) in fig. 4. We see that there is a clean signal which falls off with $p$ much the same way as the quark propagator [4]. For non-zero spatial $p$ and non-zero $q_t$ it falls off, but remains clean, while for non-zero spatial $q$ it deteriorates so rapidly that no conclusions can be drawn about its behaviour.

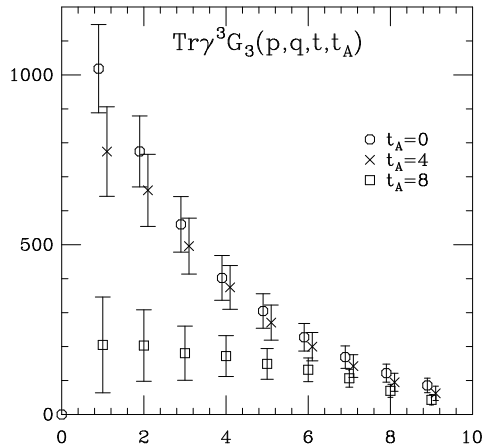

Figure 2. Zero-momentum vertex function as a function of time, for different values of the gluon time $t_A$.

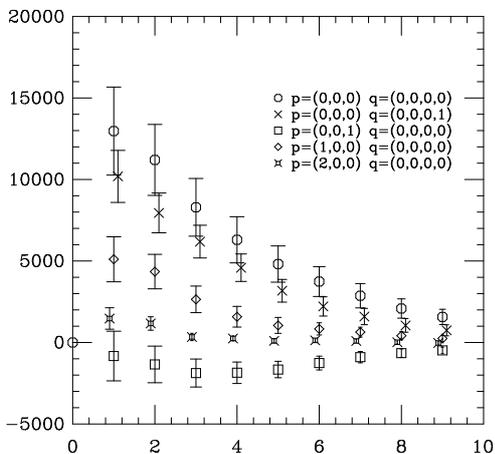

Figure 3. $\mathrm{Tr}\,\gamma^3 G_3(\vec{p},q,t)$ as a function of $t$, for different values of $\vec{p}$ and $q$.

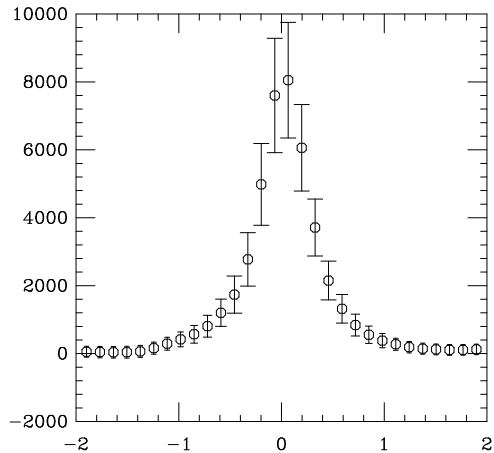

Figure 4. $\mathrm{Tr}\,\gamma^3 G_3(p,q)$ for $q=0$, $p=(\vec{p},p_t)$, as a function of $p_t a$.

It turns out that $\gamma_\mu G_\mu$ falls off smoothly with $p$ and $q$ for $p=p_\nu$ ($q=q_\nu$) with $\mu \neq \nu$. This is not the case for $\mu = \nu$; however, the signal here is too poor to make it possible to draw any conclusions about the actual behaviour of the vertex function in this case. It may be related to the $\cos(2p+q)_\mu$ term in equation (4), or to the $F_4$ and $F_7$ form factors in equations (6) and (7), but this cannot be stated with any certainty at this stage.

## 4. Discussion and conclusions

We have found a clear signal for the quark–gluon vertex in both configuration space and momentum space. However, the momentum-space vertex is still very noisy. This may be either because our statistics are too low, or because our gauge fixing is not accurate enough. Early results using gauge fixing with an accuracy of $10^{-12}$, as well as experience from study of the 3-gluon vertex [3], lend support to the latter hypothesis, but future work will involve both better gauge fixing accuracy and higher statistics.

## 5. Future work

When the uncertainties connected with the gauge fixing accuracy have been brought under control, we will go on to analyse the amputated vertex. The plan thereafter is to repeat this on a larger ensemble of configurations at $\beta = 6.2$, and possibly also at $\beta = 6.4$. This should enable us to come closer to a realistic determination of $g_R(q^2)$ from the quark–gluon vertex.

**Acknowledgements**

This work has been supported by Norwegian Research Council grant 100229/432. I wish to thank Claudio Parrinello, David Henty and David Richards for fruitful discussions.